\documentstyle[prd,aps,twocolumn,epsf,floats,amsfonts,amssymb,amsmath]{revtex}
\newcommand{\bea}{\begin{eqnarray}}\newcommand{\eea}{\end{eqnarray}}

\newcommand{\brr}{\begin{array}}\newcommand{\err}{\end{array}}
\newcommand{\bit}{\begin{itemize}}\newcommand{\eit}{\end{itemize}}
\newcommand{\ben}{\begin{enumerate}}\newcommand{\een}{\end{enumerate}}

\def\lan{\langle}\def\lar{\leftarrow}
\def\lf{\left}\def\lrar{\leftrightarrow}

\def\non{\nonumber}
\def\pa{\partial}\def\ran{\rangle}
\def\rar{\rightarrow}
\def\ri{\right}

\def\al{\alpha}\def\bt{\beta}\def\ga{\gamma}
\def\de{\delta}
\def\ep{\epsilon}\def\te{\theta}

\def\si{\sigma}
\def\om{\omega}
\def\AB{{_{A,B}}}
\begin{document}
\twocolumn[\hsize\textwidth\columnwidth\hsize\csname@twocolumnfalse\endcsname

\title{Currents and charges for mixed fields}
\author{Massimo Blasone${}^{\sharp\flat}$, Petr Jizba${}^{\natural}$
and Giuseppe Vitiello${}^{\flat }$ \vspace{3mm}}

\address{${}^{\sharp}$
Blackett  Laboratory, Imperial  College,   Prince
Consort Road, London  SW7 2BZ, U.K.
\\ [2mm] ${}^{\natural}$ Institute of Physics, University of Tsukuba,
Ibaraki 305-8571, Japan
\\ [2mm] ${}^{\flat}$
Dipartimento di  Fisica and INFN, Universit\`a di Salerno, I-84100
Salerno, Italy \vspace{2mm}}

\date{{\bf Version}, \today}

\maketitle

\begin{abstract}
We present an analysis of  currents and charges for a system of
two mixed fields, both for spinless bosons and for Dirac fermions.
This allows us to obtain in a straightforward way the exact field
theoretical
 oscillation formulas exhibiting corrections with respect to the
usual ones  derived in quantum mechanics.
\end{abstract}

\vspace{8mm} ]


\section{Introduction}

Despite being studied since long time,  particle mixing  and
oscillations \cite{cheng,neutr,zralek} are still among the most
fascinating subjects of Particle Physics: there is a renewed
interest in the physics of mixed mesons (like Kaons or B-mesons)
due to the ongoing or planned experiments \cite{mesons} and
neutrino oscillations\cite{BP78} seem finally to have been
observed \cite{Kam} after a long search. On the other hand, on the
theoretical side, many questions  remain unanswered like the ones
on the origin itself of the mixing among generations or about the
values of the lepton mixing angles\cite{Fri}. Also,  various
issues related to the phenomenology of flavor oscillations are
object of active discussions (see Refs.\cite{zralek,giunti} for
recent reviews).

Recently, there has been a remarkable progress in the
understanding of the Quantum Field Theory (QFT) of mixed
particles, for  both fermions and
bosons\cite{BV95,han00,BHV98,Fujii,remarks,lathuile,binger,BCRV99}.
It has emerged \cite{BV95,han00} that the vacuum state for the
flavor fields (describing the mixed particles) is not the same as
the one for the mass eigenstates, rather is a generalized coherent
state \cite{Per}. This fact turns out to have observable
consequences: the exact oscillation formulas obtained in QFT for
neutrinos \cite{BHV98,remarks} and for bosons \cite{BCRV99}
exhibit corrections with respect to the usual textbook expressions
derived in Quantum Mechanics (QM) \cite{cheng,neutr}.

The aim of this paper is to present an
analysis of the currents and the charges for a system of mixed
fields, which can then be used for deriving oscillation formulas.
We consider both spin zero charged bosons and Dirac fermions in
the simplest case of mixing among two generations.

The fact that such a problem has remained basically unsolved till now
is due to the  fundamental difficulties one encounters in the definition
of the Hilbert space for particles with undefined mass, as the
mixed particles are. Thus the results of Refs.\cite{BV95,BCRV99}
are crucial in the following discussion.
To our knowledge, the only existing treatment of
currents and charges for mixed particles is the one of
Ref.\cite{Anco}: the results there obtained are valid in the
framework of non-relativistic Quantum Mechanics, and therefore
have to be considered only approximate (i.e. they should be
regarded as the non-relativistic limit of exact ones). This is so
because mixing between states of different masses is not even
allowed in non-relativistic QM due to
superselection rules \cite{Barg}.

Apart from the intrinsic theoretical interest of our analysis, its
necessity has been recently emphasized in Ref.\cite{zralek}: an
unambiguous definition of currents and charges for mixed particles
would lead to a more consistent way to treat particle
oscillations, especially for the case (as it happens for actual
experiments) in which the time of measurement is not known and
what is really measured is the distance from the source. In
Ref.\cite{Anco}, oscillation formulas for the case of Kaon mixing
are obtained after integrating the probability current density
over the time of measurement and the surface of the detector ($L$
denotes the distance from the source and $\al, \bt$ are flavor
indices):
\bea\non P_{\al\rar \bt}(L) =\int_{t_1}^{t_2}dt\int_{\pa A} {\bf d
S}\cdot {\bf j}_\bt(x,t)\, , \eea
from which results in line with the usual formulas follow.

In QFT the situation is different: rather than to probability
current, one has to look at expectation values of operators
describing the flavor currents and charges. These expectation
values have to be properly defined on the Hilbert space for mixed
particles which  cannot be identified with the space for the mass
eigenstates, as shown in Refs.\cite{BV95,han00,BCRV99}. What we
show here is that currents and charges for a system of two mixed
fields can be indeed properly defined in QFT: these are the
currents associated with $SU(2)$ transformations acting on the
field doublet. By using the flavor Hilbert space
\cite{BV95,BCRV99}, we then derive  exact QFT oscillation formulas
(both for fermions and bosons) which agree with the ones already
presented in Refs.\cite{BHV98,remarks,BCRV99}. We do not discuss
here oscillation formulas in space, although our results naturally
provide a basis for such calculation, which will be treated
elsewhere.

\section{Currents and charges for mixed fields}

Let us begin our discussion with the fermion case and consider the
following Lagrangian density describing two Dirac fields with a
mixed mass term (in the following we will refer explicitly to
neutrinos):
\bea\label{lagemu} {\cal L}(x)\,=\,  {\bar \Psi_f}(x) \lf( i
\not\!\partial -
  M \ri) \Psi_f(x)\, ,
\eea
where $\Psi_f^T=(\nu_e,\nu_\mu)$ and $M = \lf(\brr{cc} m_e &
m_{e\mu} \\ m_{e\mu} & m_\mu \err \ri)$.
This Lagrangian has been already used in Ref.\cite{BHV98} to model
neutrino oscillations in vacuo.

The mixing transformations
\bea\label{fermix}
\Psi_f(x) \, =\, \lf(\brr{cc}  \cos \te & \sin \te \\
-\sin \te & \cos \te \err\ri) \Psi_m (x) \, , \eea
with $\te$ being the mixing angle and $\Psi_m^T=(\nu_1,\nu_2)$,
diagonalize the quadratic form of Eq.(\ref{lagemu}) to the
Lagrangian for two free Dirac fields, with masses $m_1$ and $m_2$:
\bea\label{lag12} {\cal L}(x)\,=\,  {\bar \Psi_m}(x) \lf( i
\not\!\partial -
  M_d\ri) \Psi_m(x)  \, ,
\eea
where $M_d = diag(m_1,m_2)$. One also has $ m_{e} =
m_{1}\cos^{2}\te + m_{2} \sin^{2}\te~$,  $m_{\mu} =
m_{1}\sin^{2}\te + m_{2} \cos^{2}\te~$,   $m_{e\mu}
=(m_{2}-m_{1})\sin\te \cos\te\,$.

The free fields  $\nu_i$ (i=1,2) can be quantized in the usual
way\cite{Itz} (we use $t\equiv x_0$):
\bea\label{2.2} \nu_{i}(x) = \sum_{r} \int d^3 k \lf[u^{r}_{{\bf
k},i} \al^{r}_{{\bf k},i}(t)\:+    v^{r}_{-{\bf k},i}\bt^{r\dag
}_{-{\bf k},i}(t)   \ri] e^{i {\bf k}\cdot{\bf x}} , \eea
with $\al^{r}_{{\bf k},i}(t)=e^{-i\om_{k,i} t}\al^{r}_{{\bf
k},i}(0)$, $\bt^{r}_{{\bf k},i}(t)=e^{-i\om_{k,i} t}\bt^{r}_{{\bf
k},i}(0)$ and  $\om_{k,i}=\sqrt{{\bf k}^2+m_i^2}$. The vacuum for
the mass eigenstates is denoted by $|0\rangle_{1,2}$:  $\; \;
\al^{r}_{{\bf k},i}|0\rangle_{1,2}= \bt^{r }_{{\bf
k},i}|0\rangle_{1,2}=0$.   The anticommutation relations are the
usual ones; the wave function orthonormality and completeness
relations are those of Ref.\cite{BV95}.

We now study the transformations acting on the doublet of free
fields with different masses. ${\cal L}$ is invariant under global
$U(1)$ phase transformations  of the type $\Psi_m' \, =\, e^{i \al
}\, \Psi_m$: as a result, we have the conservation of the Noether
charge $Q=\int d^3x \, I^0(x) $ (with $I^\mu(x)={\bar \Psi}_m(x)
\, \ga^\mu \, \Psi_m(x)$)  which is indeed the total charge of the
system (i.e. the total lepton number).

Consider now the $SU(2)$ transformations acting on
$\Psi_m$:
\bea \label{masssu2} \Psi_m'(x) \, =\, e^{i \al_j  \tau_j}\,
\Psi_m (x) \, \qquad, \qquad
 j=1, 2, 3.
\eea
with $\al_j$ real constants, $\tau_j=\si_j/2$ and $\sigma_j$ being
the Pauli matrices.

For $m_1\neq m_2$, the Lagrangian is not generally invariant under
(\ref{masssu2}) and  we obtain, by use of the equations of motion,
\bea \non &&\de {\cal L}(x)\,= \,  i \al_j \,{\bar \Psi_m}(x)\,
[\tau_j,M_d ]\, \Psi_m(x) \, =\,  - \al_j \,\pa_\mu J_{m,j}^\mu
(x)
\\ [3mm]\label{fermacu1}
&&J^\mu_{m,j}(x)\, =\,  {\bar \Psi_m}(x)\, \ga^\mu\, \tau_j\,
\Psi_m(x) \qquad, \qquad j=1, 2, 3. \eea

The charges  $Q_{m,j}(t)\equiv \int d^3 x \,J^0_{m,j}(x) $,
satisfy the  $su(2)$  algebra $[Q_{m,j}(t), Q_{m,k}(t)]\, =\, i
\,\ep_{jkl} \,Q_{m,l}(t) $. Note that the Casimir operator is
proportional to the total charge: $C_m \equiv \Big[
\sum\limits_{j=1}^3 Q_{m,j}^2(t)\Big]^{\frac{1}{2}}$
$=\,\frac{1}{2}Q$. From (\ref{fermacu1}) we also see  that
$Q_{m,3}$ is conserved as $M_d$ is diagonal. We can define the
combinations:
\bea\label{noether1} &&Q_{1}\, \equiv \,\frac{1}{2}Q \,+ \,Q_{m,3}
\quad, \quad Q_{2}\, \equiv \,\frac{1}{2}Q \,- \,Q_{m,3}
\\ \non
&&Q_i \, = \,\sum_{r} \int d^3 k\lf( \al^{r\dag}_{{\bf k},i}
\al^{r}_{{\bf k},i}\, -\, \bt^{r\dag}_{-{\bf k},i}\bt^{r}_{-{\bf
k},i}\ri) \quad, \quad i=1, 2. \eea

These are nothing but  the Noether charges associated with the
non-interacting fields $\nu_1$  and $\nu_2$: in the absence of
mixing, they are the flavor charges,  separately
conserved for each generation.

Observe now that the transformation:
\bea \Psi_f(x) &=& e^{-2 i  \te Q_{m,2}(t) } \Psi_m(x) e^{2 i \te
Q_{m,2}(t) } \eea
is just the mixing Eq.(\ref{fermix}). Thus $ 2 Q_{m,2}(t)$
is the generator of the mixing transformations:
its properties have been studied in ref.\cite{BV95}, where it has
been shown that the action  of $G_\te(t) \equiv e^{2 i
\te Q_{m,2}(t)}$ on the vacuum state $|0\ran_{1,2}$
results in a new vector (flavor vacuum) $|0(t)\ran_{e,\mu}\equiv
G^{-1}_\te(t)|0\ran_{1,2}$, orthogonal to $|0\ran_{1,2}$ in the
infinite volume limit.

By use of $G_\te(t)$, the flavor fields can be expanded as:
\bea \non \label{exnue1}
\nu_\si(x)&=& \sum_{r} \int d^3 k   \lf[ u^{r}_{{\bf k},i}
\al^{r}_{{\bf k},\si}(t) +    v^{r}_{-{\bf k},i}
\bt^{r\dag}_{-{\bf k},\si}(t) \ri]  e^{i {\bf k}\cdot{\bf x}}\,.
\eea
with $(\si,i)=(e,1) , (\mu,2)$.
The flavor annihilation operators are defined as $\al^{r}_{{\bf
k},\si}(t) \equiv G^{-1}_{\te}(t)\al^{r}_{{\bf k},i}(t)
G_{\te}(t)$ and $\bt^{r\dag}_{{-\bf k},\si}(t)\equiv
 G^{-1}_{\te}(t)  \bt^{r\dag}_{{-\bf k},i}(t)
G_{\te}(t)$.

Let us now perform the  $SU(2)$ transformations on the flavor
doublet $\Psi_f$:
\bea \Psi_f'(x) \, =\, e^{i  \al_j \tau_j }\, \Psi_f (x) \qquad
,\qquad j \,=\, 1, 2, 3. \eea
We obtain
\bea \non &&\de {\cal L}(x)\,= \,   i \al_j\,{\bar \Psi_f}(x)\,
[\tau_j, M]\, \Psi_f(x)\, =\, - \al_j \,\pa_\mu J_{f,j}^{\mu}(x)\,
,
\\ [3mm] \label{fermacu2}
&&J^\mu_{f,j}(x) \,=\,  {\bar \Psi_f}(x)\, \ga^\mu\, \tau_j\,
\Psi_f(x)\qquad ,\qquad j \,=\, 1, 2, 3. \eea

The related charges $Q_{f,j}(t)$ $\equiv$  $\int d^3 x
\,J^0_{f,j}(x) $, still close the $su(2)$ algebra and $C_f=C_m
=\frac{1}{2}Q$. Due to the off--diagonal (mixing) terms in the
mass matrix $M$, $Q_{f,3}(t)$ is  time--dependent. This implies an
exchange of charge between $\nu_e$ and $\nu_\mu$, resulting in the
flavor oscillations.

In accordance with Eq.(\ref{noether1}), we define the {\em flavor
charges} for mixed fields as
\bea Q_e(t) & \equiv & \frac{1}{2}Q \, + \, Q_{f,3}(t)
\\
Q_\mu(t) & \equiv & \frac{1}{2}Q \, -  \, Q_{f,3}(t) \eea
with $Q_e(t) \, + \,Q_\mu(t) \, = \, Q$.

These charges have a simple expression in terms of the flavor
ladder operators ($\si= e,\mu$):
\bea\non Q_\si(t) & = & \sum_{r} \int d^3 k\lf( \al^{r\dag}_{{\bf
k},\si}(t) \al^{r}_{{\bf k},\si}(t)\, -\, \bt^{r\dag}_{-{\bf
k},\si}(t)\bt^{r}_{-{\bf k},\si}(t)\ri)\,, \eea
which happens  because they are connected to the Noether charges
$Q_i$
 of Eq.(\ref{noether1}) via the mixing generator: $Q_\si(t)
= G^{-1}_{\te}(t)Q_i G_{\te}(t)$. Notice also that the operator
\bea \Delta Q_e&\equiv & Q_e(t) - Q_1 \, =\, Q_{f,3}(t) - Q_{m,3}
\eea
describes how much the mixing violates the lepton number
conservation for a given generation.

 The oscillation formulas are
obtained by taking expectation values of the above charges on the
(flavor) neutrino state. Consider for example an initial electron
neutrino state defined as $|\nu_e\ran \equiv \al_{{\bf k},e}^{r
\dag}(0) |0\ran_{e,\mu}$ (for a discussion on the  correct
definition of flavor states see Refs.\cite{BHV98,remarks,BCRV99}).
Working in the Heisenberg picture, we obtain
\bea \label{charge1} &&\;_{e,\mu}\langle 0|Q_\si(t)|
0\rangle_{e,\mu} \,= \, 0 \, ,
\\[2mm] \label{charge2}
&&{\cal Q}_{{\bf k},\si}(t) \,\equiv\, \langle \nu_e|Q_\si(t)|
\nu_e\rangle
\\ \non
&&\;=\, \lf|\lf \{\al^{r}_{{\bf k},\si}(t), \al^{r \dag}_{{\bf
k},e}(0) \ri\}\ri|^{2} \;+ \;\lf|\lf\{\bt_{{-\bf k},\si}^{r
\dag}(t), \al^{r \dag}_{{\bf k},e}(0) \ri\}\ri|^{2} \, . \eea
where $|0\ran_{e,\mu}\equiv |0(0)\ran_{e,\mu}$. Charge
conservation is obviously ensured at any time: ${\cal Q}_{{\bf
k},e}(t) + {\cal Q}_{{\bf k},\mu}(t)\; = \; 1$. We remark that the
expectation value of $Q_\si$ cannot be taken on vectors of the
Fock space built on $|0\rangle_{1,2}$, as shown in
Refs.\cite{BHV98,remarks,BCRV99}.

The oscillation (in time) formulas for the flavor charges, on an
electron neutrino state, then follow:
\bea\non {\cal Q}_{{\bf k},e}(t) \, = \,1 &-& \sin^{2}( 2 \te) \,
|U_{{\bf k}}|^{2}\; \sin^{2} \lf( \frac{\om_{k,2} - \om_{k,1}}{2}
t \ri)
\\ \label{oscillferm}
&-& \sin^{2}( 2 \te)\,|V_{{\bf k}}|^{2} \; \sin^{2} \lf(
\frac{\om_{k,2} + \om_{k,1}}{2} t \ri) \, ,
\end{eqnarray}
%
with $|U_{{\bf k}}|^{2}+|V_{{\bf k}}|^{2}=1$ and
\bea &&|V_{{\bf k}}|=\frac{|{\bf k}|}{2\sqrt{\om_{k,1}\om_{k,2}}}
\lf(\sqrt{\frac{\om_{k,1}+m_{1}}{\om_{k,2}+m_{2}}}-
\sqrt{\frac{\om_{k,2}+m_{2}}{\om_{k,1}+m_{1}}}\ri) \eea

This result is exact. The differences with respect to the usual
formula for neutrino oscillations are in the energy dependence of
the amplitudes and in the additional oscillating term.
For $|{\bf k}|\gg\sqrt{m_1m_2}$, we have $|U_{{\bf
k}}|^{2}\rar 1$ and  $|V_{{\bf k}}|^{2}\rar 0$ and the traditional
(Pontecorvo) oscillation formula is recovered.

Finally, it is useful to define the {\em flavor currents} in the
following way:
\bea \label{fcurr1} J^\al_e(x) & \equiv & \frac{1}{2} I^\al (x) \,
+ \, J_{f,3}(x)\, ,
\\  \label{fcurr2}
J^\al_\mu(x) & \equiv & \frac{1}{2} I^\al (x) \, - \, J_{f,3}(x)\,
. \eea
The sum of these two currents gives $I^\al$, i.e. the Noether
current density for the total system. These quantities are the
natural candidates for a QFT formulation of flavor oscillations in
space.

\vspace{0.5cm}

For the  bosonic case we can proceed in a similar way. For
simplicity, we deal with spin zero stable particles. The
Lagrangian for two  mixed charged scalar fields is (where necessary,
we use a ``hat''
symbol  to distinguish similar quantities from the fermionic
case):
\bea\label{boslagAB} {\cal L}(x)&=& \pa_\mu \Phi_f^\dag(x) \pa^\mu
\Phi_f (x)\, - \, \Phi_f^\dag(x) {\hat M}  \Phi_f(x)
\\ \label{boslag12}
&=&\pa_\mu \Phi_m^\dag(x) \pa^\mu \Phi_m(x) \, - \, \Phi_m^\dag(x)
{\hat M}_d  \Phi_m (x) \eea
with $\Phi_f^T=(\phi_A,\phi_B)$ being the flavor fields and ${\hat
M} = \lf(\brr{cc} m_A^2 & m_{AB}^2 \\ m_{AB}^2 & m_B^2\err \ri)$.
Those are connected to the  free fields $\Phi_m^T=(\phi_1,\phi_2)$
with  ${\hat M}_d = diag(m_1^2,m_2^2)$ by a rotation similar to
Eq.(\ref{fermix}). For an extensive treatment of boson mixing in
QFT, see Ref.\cite{BCRV99}.

Similarly to the fermionic case, we consider the transformation
$\Phi_m'(x) = e^{i \al_j \tau_j}\, \Phi_m(x) $ with $j= 1,2,3$,
which gives
\bea \non &&\de {\cal L}(x)\,= \,   i\,\al_j\, \Phi_m^\dag(x) \,
[\tau_j , {\hat M}_d] \, \Phi_m(x)\, = \,- \al_j \,\pa_\mu\,
{\hat J}^\mu_j(x)
\\
&&{\hat J}^\mu_{m,j}(x) \,=\, i \, \Phi_m^\dag(x) \, \tau_j\,
\stackrel{\lrar}{\pa^\mu}\, \Phi_m (x) \quad ,\quad j \,=\, 1, 2,
3 , \eea
where $ \stackrel{\lrar}{\pa^\mu} \, \equiv\,
\stackrel{\rar}{\pa^\mu} \! -\! \stackrel{\lar}{\pa^\mu} $.

The charges ${\hat Q}_{m,i}(t)\equiv $ $\int d^3 x \,{\hat
J}^0_{m,j}(x) $,  satisfy the $su(2)$ algebra (at each time $t$).
$2  {\hat Q}_{m,2}(t)$ is the generator of the boson mixing
transformations whose properties are studied in
Refs.\cite{lathuile,binger,BCRV99}. The ``flavor'' vacuum is
defined as $|0(t)\ran_\AB \equiv {\hat G}^{-1}_\te(t)|{\hat
0}\ran_{1,2}$ with ${\hat G}_\te(t) \equiv e^{i 2 \te {\hat
Q}_{m,2}(t) }$ and the flavor annihilation operators are $a_{{\bf
k},A}(t) \equiv {\hat G}_\te^{-1}(t) \; a_{{\bf k},1}\;{\hat
G}_\te(t)$, etc.. As in the fermionic case, these operators
involve Bogoliubov coefficients, which now satisfy the hyperbolic
relation $|{\hat U}_{{\bf k}}|^{2} - |{\hat V}_{{\bf k}}|^{2}=1 $
with \cite{lathuile,binger,BCRV99} \bea |{\hat V}_{{\bf k}}|\equiv
\frac{1}{2} \lf( \sqrt{\frac{\om_{k,1}}{\om_{k,2}}}
-\sqrt{\frac{\om_{k,2}}{\om_{k,1}}}  \ri) \, . \eea
The flavor charges for bosons are:
\bea\non {\hat Q}_\si(t) & = & \int d^3 k \lf( a^{\dag}_{{\bf
k},\si}(t) a_{{\bf k},\si}(t)\, -\, b^{\dag}_{-{\bf
k},\si}(t)b_{-{\bf k},\si}(t)\ri)\,, \eea
where $\si=A,B$. We define a flavor $``A"$ state as
$|a_A\ran\equiv a_{{\bf k},A}^\dag(0) |0\ran_\AB$. On this state,
the expectation values of the flavor charges give
\bea {\hat {\cal Q}}_\si(t)&\equiv& \lan a_A | {\hat Q}_\si(t) |
a_A\ran
\\ \non
&=& \lf|\lf[a_{{\bf k},\si}(t), a^{\dag}_{{\bf k},A}(0) \ri]\ri|^2
\; - \; \lf|\lf[b^\dag_{{\bf k},\si}(t), a^{\dag}_{{\bf k},A}(0)
\ri]\ri|^2  \, . \eea
We also have $\,_\AB\langle 0|{\hat Q}_\si(t)| 0\rangle_\AB =0$
and ${\hat {\cal Q}}_{{\bf k},A}(t) + {\hat {\cal Q}}_{{\bf
k},B}(t)=  1$. The explicit calculation  gives
\bea \non {\hat {\cal Q}}_{{\bf k},A}(t) \,=\,
 1 &-& \sin^{2}( 2 \te) \, |{\hat U}_{{\bf k}}|^{2} \;
\sin^{2} \lf( \frac{\om_{k,2} - \om_{k,1}}{2} t \ri)
\\ \label{oscillbos}
&+&\sin^{2}( 2 \te) \,|{\hat V}_{{\bf k}}|^{2} \; \sin^{2} \lf(
\frac{\om_{k,2} + \om_{k,1}}{2} t \ri) \, .
\eea
For $|{\bf k}|^2\gg \frac{m_1^2 +m_2^2}{2}$
we have $|{\hat V}_{{\bf k}}| \rar  0$ and the usual result is recovered.
Notice the  negative sign in front of $|{\hat V}_{{\bf k}}|^{2}$ in
contrast with the one for fermions: the oscillating flavor charge
for bosons can assume also negative values (of course,  charge
conservation holds for the overall system of two mixed fields).
This happens since in
QFT, flavor states are essentially multiparticle ones and one has
to abandon the single--particle
probabilistic interpretation of QM in favor of a
statistical picture.

It is important to stress again that the use of the flavor Hilbert
space, i.e. of the Hilbert space constructed on the flavor vacuum,
is absolutely crucial in obtaining the above oscillation formulas
Eqs.(\ref{oscillferm}), (\ref{oscillbos}). In fact, although the
currents and the charges for the flavor fields are defined purely
from  symmetry considerations on the Lagrangian, the choice of the
representation is essential when their expectation values have to
be defined. It turns out indeed\cite{Fujii,remarks,BCRV99}, that
arbitrary mass parameters can be introduced in the expansion of
the flavor fields, and that the only quantities free from such
arbitrary quantities are the expectation  values of the flavor
charges on states defined as above in the flavor Hilbert space.

\section{Conclusions}

In this paper, we have shown that it is possible to define
consistently  currents and charges for a system of mixed fields,
i.e. for fields with no definite masses. This was done in the case
of two flavors, by studying the $SU(2)$ transformations acting on
a doublet of fields with different masses: in this way, we
naturally recovered the generator of the mixing transformations
already introduced in Refs.\cite{BV95,BCRV99} for fermions and
bosons, respectively. Our analysis also leads in a straightforward
way to the exact QFT oscillation formulas found recently
\cite{BHV98,remarks,BCRV99} which exhibit relevant corrections
with respect to the usual textbook QM ones.

Our results provide a basis for an exact QFT calculation of
oscillation formulas with dependence on space rather than time, a
situation which is much closer to the experimental setup. This is
one of the aspect which we intend to study in the future.

In the present paper we considered two--flavor mixing for
simplicity. The above analysis can be extended to the mixing among
an higher number of generations. The case of three--flavor fermion
mixing is particularly relevant for the phenomenology (quarks,
neutrinos) and such an extension is currently under investigation.

\vspace{0.5cm}
This work has been partially supported by INFN and MURST.


\end{document}